# STATISTICAL ANALYSIS AND INFORMATION THEORY OF SCREENED KRATZER-HELLMANN POTENTIAL MODEL

G. T. Osobonye[1], U. S. Okorie[2 & 3*], P. O. Amadi[2], A. N. Ikot[2]

[1]Department of Physics, Federal College of Education (Technical), Omoku, Rivers State, Nigeria.
[2]Department of Physics, University of Port Harcourt, Choba, Rivers State, Nigeria.
[3]Department of Physics, Akwa Ibom State University, Ikot Akpaden, Akwa Ibom State, Nigeria.

**Abstract**
In this research, the radial Schrodinger equation for a newly proposed screened Kratzer-Hellmann potential model was studied via the conventional Nikiforov-Uvarov method. The approximate bound state solution of the Schrodinger equation was obtained using the Greene-Aldrich approximation, in addition to the normalized eigenfunction for the new potential model both analytically and numerically. These results were employed to evaluate the rotational-vibrational partition function and other thermodynamic properties for the screened Kratzer-Hellmann potential. We have discussed the results obtained graphically. Also, the normalized eigenfunction has been used to calculate some information-theoretic measures including Shannon entropy and Fisher information for low lying states in both position and momentum spaces numerically. We observed that the Shannon entropy results agreed with the Bialynicki-Birula and Mycielski inequality, while the Fisher information results obtained agreed with the Stam, Crammer-Rao inequality. From our results, we observed alternating increasing and decreasing localization across the screening parameter in the both eigenstates.

**Corresponding Author's Email:** uduakobongokorie@aksu.edu.ng

## 1. Introduction

Many years ago, a researcher called Kratzer [1] established a potential that has been widely studied and applied in different fields. This potential has contributed immensely to the description of molecular structure both in non-relativistic and relativistic mechanical systems [2, 3]. Prominently, the Kratzer potential is employed in the study of atomic physics, molecular physics and quantum chemistry [4]. An exact solution of the Schrodinger equation for Kratzer potential was obtained by Bayrak et al. [5], via the asymptotic iteration method (AIM). Researchers have also modified this potential for broader applications. Sadeghi [6] studied the noncentral modified Kratzer potential with the Schrodinger equation using the factorization method. An approximate solution of the Schrodinger equation for the modified Kratzer plus screened Coulomb potential model has been obtained for diatomic molecules [7]. Of recent, a screened Kratzer potential has been proposed [8], which has been employed to study expectation values of Lithium hydride and hydrogen chloride diatomic molecules.

The Hellmann potential [9] is known to be a combination of Coulomb and Yukawa potentials. Different authors have reported the various studies carried out with the Hellmann potential [10-15].

Having these lofty applications mentioned above, we are motivated to propose the screened Kratzer-Hellmann potential (SKHP) defined to be

$$V(r) = \left( \frac{V_0}{r} + \frac{V_1}{r} e^{\alpha r} + \frac{V_2}{r^2} \right) e^{-\alpha r} \qquad (1),$$

where $V_0$, $V_1$ and $V_2$ are potential parameters and $\alpha$ is the screening parameter. It can be deduced that the SKHP reduces to the Hellmann potential when $V_2 = 0$; the screened Kratzer potential when $V_1 = 0$; Kratzer potential when $V_1 = \alpha = 0$; screened Coulomb (or Yukawa) potential when $V_1 = V_2 = 0$; and Coulomb potential when $V_1 = V_2 = \alpha = 0$. Hence, the SKHP is a generalized potential composed of the Hellmann, screened Kratzer, Kratzer, Yukawa and Coulomb potentials, respectively.

By employing the rotation-vibrational energies obtained for various potential models, the partition function and other thermodynamic functions of various molecules have been studied by different authors [16-22]. Recently, Ikot and his collaborators [8] studied the screened Kratzer potential in the non-relativistic regime and obtained its vibrational partition function and other thermodynamic properties of lithium hydride and hydrogen chloride diatomic molecules in closed form.

The concept of information theory and its application to quantum mechanical systems has attracted the attention of many researchers since its foundation was laid by C. E. Shannon [23]. Since then, the information theory found its way into communication theory [24-26]. The Shannon entropy find is significance in density functional theory [27], the study of chemical properties of atomic, molecular and reactive systems [28-30]. Information-theoretic measures study the probability distribution across quantum mechanical states in position and momentum [31].

In recent developments, quantum information entropies have been evaluated for different potential models [32-34]. Najafizade et al. [35] studied the Shannon information entropy of Kratzer potential within the nonrelativistic framework. Onate et al. [36] evaluated the Tsallis and Renyi entropies in position and momentum spaces under the Hellmann potential. The information entropies of modified Hulthen potential (MHP) in the position and momentum spaces have been calculated using the J-matrix method [37]. By employing the confined isotropic harmonic oscillator, Makherjee and Roy [38] studied the information-based uncertainty measures including the Renyi entropy, Shannon entropy and Onicescu energy. Also, Isonguyo et al. [39] studied the quantum information-theoretic analysis of the static screened Coulomb potential, both analytically and numerically. Most recently, Onate and his collaborators [40] investigated the Fisher information and uncertainty relation of a special potential family. Up till now, the screened Kratzer - Hellmann potential has not been studied within the framework of information theory; hence our motivation.

The research aims to investigate the statistical functions and information-theoretic measures of the screened Kratzer-Hellmann potential model. This article is organized as follows: In section 2, the eigenvalues and eigenfunctions of the Schrodinger equation are obtained with screened SKHP. Some special cases of the SKHP are obtained in section 3. In section 4, the rotational-vibrational partition function and other thermodynamic properties of the SKHP are obtained using the energy eigenvalues earlier derived. In section 5, we evaluate some information-theoretic measures of the SKHP, with the help of the eigenfunctions obtained Shannon entropy and Fisher information is studied numerically in three dimension. The entire results obtained are discussed extensively in section 6. Finally, the concluding remarks of this work are given in section 7.

## 2. Eigensolutions of Schrodinger equation with screened Kratzer-Hellmann potential (SKHP)

The radial part of the Schrödinger equation is given by [41];

$$\frac{d^2\psi_{n\ell}(r)}{dr^2} + \frac{2\mu}{\hbar^2}\left[E_{n\ell} - V(r) - \frac{\hbar^2\ell(\ell+1)}{2\mu r^2}\right]\psi_{n\ell}(r) = 0 \tag{2}$$

$\mu$ is the reduced mass, $E_{nl}$ is the energy spectrum to be determined, $\hbar$ is the reduced Planck's constant and $n$ $and$ $\ell$ are the radial and orbital angular momentum quantum numbers respectively.

Substituting Eq. (1) into Eq. (2) gives

$$\frac{d^2\psi_{n\ell}(r)}{dr^2} + \left[\frac{2\mu E_{nl}}{\hbar^2} - \frac{2\mu}{\hbar^2}\left(\frac{V_0}{r} + \frac{V_1}{r}e^{\alpha r} + \frac{V_2}{r^2}\right)e^{-\alpha r} - \frac{\ell(\ell+1)}{r^2}\right]\psi_{n\ell}(r) = 0 \tag{3}$$

By employing both the approximation Scheme [42]:

$$\frac{1}{r^2} \approx \frac{\alpha^2}{\left(1-e^{-\alpha r}\right)^2}, \quad \frac{1}{r} \approx \frac{\alpha}{\left(1-e^{-\alpha r}\right)} \tag{4},$$

and a coordinate transformation of the form $z = e^{-\alpha r}$, Eq. (3) reduces to the differential equation given as

$$\frac{d^2\psi_{n\ell}}{dz^2} + \frac{(1-z)}{z(1-z)}\frac{d\psi_{n\ell}}{dz} + \frac{1}{z^2(1-z)^2}\left[-\left(\varepsilon^2 + A\right)z^2 + \left(B + 2\varepsilon^2\right)z + \left(C - \varepsilon^2\right)\right]\psi_{n\ell}(z) = 0 \tag{5},$$

where

$$\varepsilon^2 = -\frac{2\mu E_{n\ell}}{\hbar^2\alpha^2}; \quad A = -\frac{2\mu V_0}{\hbar^2\alpha}; \quad B = \left(\frac{2\mu V_1}{\hbar^2\alpha} - \frac{2\mu V_2}{\hbar^2} - \frac{2\mu V_0}{\hbar^2\alpha}\right); \quad C = -\left[\frac{2\mu V_1}{\hbar^2\alpha} + \ell(\ell+1)\right] \tag{6}.$$

By invoking the NU method (see appendix A), we obtain the non-relativistic energy eigenvalues of the SKHP and its corresponding unnormalized eigenfunction to be

$$E_{n\ell} = \alpha V_1 + \frac{\hbar^2\alpha^2\ell(\ell+1)}{2\mu} - \frac{\hbar^2\alpha^2}{2\mu}\left[\frac{-\left(\frac{2\mu V_0}{\hbar^2\alpha} + \frac{2\mu V_1}{\hbar^2\alpha} + \ell(\ell+1)\right) - \left(n + \frac{1}{2} + \sqrt{\frac{2\mu V_2}{\hbar^2} + \ell(\ell+1) + \frac{1}{4}}\right)^2}{2\left(n + \frac{1}{2} + \sqrt{\frac{2\mu V_2}{\hbar^2} + \ell(\ell+1) + \frac{1}{4}}\right)}\right]^2 \tag{7},$$

$$\psi_{n\ell}(z) = N_{n\ell}\, z^{\sqrt{\varepsilon^2 - C}}(1-z)^G P_n^{\left(2\sqrt{\varepsilon^2 - C},\, 2G-1\right)}(1-2z) \tag{8},$$

where

$$G = \frac{1}{2}\left(1 + \sqrt{1 + 4(A - B - C)}\right) \tag{9}.$$

We also find the normalization constant by writing the radial wave function as

$$\left[\psi(z)\right]^2 = N_{n\ell}^2\, z^{2\sqrt{\varepsilon^2 - C}}(1-z)^{2G}\left[P_n^{\left(2\sqrt{\varepsilon^2 - C},\, 2G-1\right)}(1-2z)\right]^2, \quad z = e^{-\alpha r} \tag{10}.$$

Eq. (10) also represents the probability density $\rho(s)$. By employing the normalization condition given by

$$\int_0^\infty \left|\psi(r)\right|^2 dr = 1 \tag{11},$$

$$-\frac{N_{n\ell}^2}{\alpha}\int\limits_1^0 z^{2\sqrt{\varepsilon^2-C}}(1-z)^{2G}\left[P_n^{\left(2\sqrt{\varepsilon^2-C},\,2G-1\right)}(1-2z)\right]^2\frac{dz}{z}=1,\ z=e^{-\alpha r} \tag{12}.$$

We carry out a coordinate transformation $S=1-2z$ in Eq. (12) to have

$$\frac{N_{n\ell}^2}{2\alpha}\int\limits_{-1}^1\left(\frac{1-S}{2}\right)^{2\sqrt{\varepsilon^2-C}}\left(\frac{1+S}{2}\right)^{2G}\left[P_n^{\left(2\sqrt{\varepsilon^2-C},\,2G-1\right)}(S)\right]^2 dS=1 \tag{13}.$$

Using the standard integral [43],

$$\int\limits_{-1}^1\left(\frac{1-t}{2}\right)^a\left(\frac{1+t}{2}\right)^b\left[P_n^{(a,b-1)}(t)\right]^2 dt=\frac{2^{a+b+1}\Gamma\left(a+n+1\right)\Gamma\left(b+n+1\right)}{n!\,\Gamma\left(a+b+n+1\right)\Gamma\left(a+b+2n+1\right)} \tag{14},$$

We obtain the normalization constant in eq. (13) as

$$N_{n\ell}=\sqrt{\frac{2\alpha\left(n!\right)\Gamma\left(1+n+2G+2\sqrt{\varepsilon^2-C}\right)\Gamma\left(1+2n+2G+2\sqrt{\varepsilon^2-C}\right)}{2^{\left(1+2G+2\sqrt{\varepsilon^2-C}\right)}\Gamma\left(1+n+2\sqrt{\varepsilon^2-C}\right)\Gamma\left(1+n+2G\right)}} \tag{15}.$$

### 3. Special Cases of Screened Kratzer-Hellmann Potential (SKHP)

Here, we consider specific special cases of interest.

- If the potential parameter $V_1$ reduces to zero, Eq. (1) also reduces to the screened Kratzer potential given as

$$V(r)=\left(\frac{V_0}{r}+\frac{V_2}{r^2}\right)e^{-\alpha r} \tag{16}.$$

Its corresponding energy eigenvalue relation is obtained to be

$$E_{n\ell}=\frac{\hbar^2\alpha^2\ell\left(\ell+1\right)}{2\mu}-\frac{\hbar^2\alpha^2}{2\mu}\left[\frac{-\left(\frac{2\mu V_0}{\hbar^2\alpha}+\ell\left(\ell+1\right)\right)-\left(n+\frac{1}{2}+\sqrt{\frac{2\mu V_2}{\hbar^2}+\ell\left(\ell+1\right)+\frac{1}{4}}\right)^2}{2\left(n+\frac{1}{2}+\sqrt{\frac{2\mu V_2}{\hbar^2}+\ell\left(\ell+1\right)+\frac{1}{4}}\right)}\right]^2 \tag{17}.$$

Eq. (17) is consistent with the result obtained in Eq. (29) of ref. [8].

- Setting the potential parameter $V_2$ equal to zero, Eq. (1) reduces to the Hellmann potential given as

$$V(r) = \frac{V_0}{r} e^{-\alpha r} + \frac{V_1}{r} \tag{18}$$

The energy eigenvalue relation of the Hellmann potential resulting from Eq. (18) is obtained as

$$E_{n\ell} = \alpha V_1 + \frac{\hbar^2 \alpha^2 \ell(\ell+1)}{2\mu} - \frac{\hbar^2 \alpha^2}{2\mu} \left[ \frac{-\left(\frac{2\mu V_0}{\hbar^2 \alpha} + \frac{2\mu V_1}{\hbar^2 \alpha} + \ell(\ell+1)\right) - \left(n + \frac{1}{2} + \sqrt{\ell(\ell+1) + \frac{1}{4}}\right)^2}{2\left(n + \frac{1}{2} + \sqrt{\ell(\ell+1) + \frac{1}{4}}\right)} \right]^2 \tag{19}$$

Eq. (19) is consistent with the result obtained in Eq. (21) of ref. [44].

### 4. Statistical Properties of Screened Kratzer-Hellmann Potential (SKHP)

The bound state contributions to the rotation-vibrational partition function of any system at a given temperature T is given as [45]

$$Z(\beta, \lambda) = \sum_{n=0}^{\lambda} e^{-\beta E_{n\ell}}, \ \beta = (k_B T)^{-1} \tag{20}$$

where $k_B$ is the Boltzmann's constant, $\lambda$ is the upper bound quantum number, $E_{n\ell}$ is the rotation-vibrational energy eigenvalues of the SKHP.

Substituting Eq. (7) into Eq. (20), we obtain

$$Z(\beta, \lambda) = \sum_{n=0}^{\lambda} e^{-\beta\left[P_2 - \frac{\hbar^2 \alpha^2}{2\mu}\left(\frac{P_1}{2(n+\sigma)} - \frac{(n+\sigma)}{2}\right)^2\right]} \tag{21}$$

where

$$P_1 = -\left[\frac{2\mu}{\hbar^2 \alpha}(V_0 + V_1) + \ell(\ell+1)\right], \ P_2 = \alpha V_1 + \frac{\hbar^2 \alpha^2}{2\mu}\ell(\ell+1), \ \sigma = \frac{1}{2}\left(1 + \sqrt{(1+2\ell)^2 + \frac{8\mu V_2}{\hbar^2}}\right) \tag{22}$$

Replacing the sum by an integral in the classical limit, we obtain:

$$Z = \int_{\sigma}^{\lambda+\sigma} e^{\left[H_1 \beta \rho^2 + \frac{H_2}{\rho^2}\beta - H_3 \beta\right]} d\rho, \ \rho = n + \sigma \tag{23}$$

where

$$H_1 = \frac{\hbar^2 \alpha^2}{8\mu}; \quad H_2 = \frac{\hbar^2 \alpha^2 P_1^2}{8\mu}; \quad H_3 = \frac{\hbar^2 \alpha^2 P_1}{4\mu} + P_2 \tag{24}.$$

By employing either a Maple or a Mathematica software to evaluate the integral in Equation (23), we obtain the rotation-vibrational partition function of the SKHP to be

$$Z\left(\beta, \lambda\right) = e^{\beta\left(H_1 \rho^2 - H_3\right)} \left[ \sqrt{H_2 \beta} \sqrt{\pi} \left( erfi\left( \frac{\sqrt{H_2 \beta}}{(\lambda + \sigma)} \right) - erfi\left( \frac{\sqrt{H_2 \beta}}{\sigma} \right) \right) + \sigma e^{\frac{H_2 \beta}{\sigma^2}} - \left(\sigma + \lambda\right) e^{\frac{H_2 \beta}{(\lambda + \sigma)^2}} \right]$$

(25).

Here, the imaginary error function is defined as [45]

$$erfi(u) = \frac{erf(i\,u)}{i} = \frac{2}{\sqrt{\pi}} \int_0^u e^{s^2} ds \tag{26}.$$

By employing Eq. (26), we can obtain other thermodynamic properties of the SKHP with the following relations:

➢ Ro-vibrational Internal energy

$$U\left(\beta, \lambda\right) = -\frac{\partial \ln Z\left(\beta, \lambda\right)}{\partial \beta} \tag{27}$$

➢ Ro-vibrational free energy

$$F\left(\beta, \lambda\right) = -\frac{1}{\beta} \ln Z\left(\beta, \lambda\right) \tag{28}$$

➢ Ro-vibrational entropy

$$S\left(\beta, \lambda\right) = k_B \ln Z\left(\beta, \lambda\right) - k_B \beta \frac{\partial \ln Z\left(\beta, \lambda\right)}{\partial \beta} \tag{29}$$

➢ Ro-vibrational specific heat capacity

$$C\left(\beta, \lambda\right) = k_B \beta^2 \frac{\partial^2}{\partial \beta^2} \ln Z\left(\beta, \lambda\right) \tag{30}$$

### 5.  Information-Theoretic Measures of Screened Kratzer-Hellmann Potential (SKHP)

It has been established that among the global information-theoretical tools, the Shannon entropy is fundamental in the uncertain measure of probability distribution of particle localization [46]. Based on position and momentum space entropies, Bialynicki-Birula and Mycielski (BBM) [47] derived the entropic relation for Shannon entropy. The uncertainty relations is better than Heisenberg uncertainty relation as it can account for higher other. The Shannon entropy relation is better than is given as [47],

$$S_r + S_p \geq D\left(1 + \ln \pi\right) \tag{31},$$

where $S_r$ and $S_p$ are the position space and momentum space Shannon entropies, respectively and $D$ represent the spatial dimension. The Shannon entropies for both the position and momentum spaces are expressed respectively in one dimension as [44]

$$S_r = -\int \rho(r) Log\left[\rho(r)\right] dr \tag{32},$$

$$S_p = -\int \rho(p) Log\left[\rho(p)\right] dr \tag{33}.$$

Here, $\rho(r) = \left|\psi(r)\right|^2$ represents the probability density in the position space, $\rho(p) = \left|\psi(p)\right|^2$ is the probability density in the momentum space and $\psi(p)$ is the wave function in the momentum space, which is obtained by carrying out the Fourier transform of the wave function in the position space, $\psi(r)$. It is imperative to note that the foundation of information theory lies in the probability density function which can be obtained in both position and momentum spaces. Due to the rigorous process in obtaining these wave functions at higher states, we restrict our calculations to the low lying states, $n = 0$ and $n = 1$ numerically.

The Fisher Information is a local measure information-theoretic measure of the probability distribution. It provides an insight into the density function of the probability density [39]. The Fisher information detect changes of probability density. In position and momentum spaces, it is expressed as follows:

$$I_r = \int \frac{\left[\rho'(r)\right]^2}{\rho(r)} dr = 4 \int \left|\psi'(r)\right|^2 dr = 4\left\langle p^2 \right\rangle - 2(2l+1)|m|\left\langle r^{-2} \right\rangle \tag{34}$$

$$I_p = \int \frac{\left[\rho'(p)\right]^2}{\rho(p)} dp = 4 \int \left|\psi'(p)\right|^2 dp = 4\left\langle r^2 \right\rangle - 2(2l+1)|m|\left\langle p^{-2} \right\rangle \tag{35},$$

The Fisher information product can as well be expressed as [49,50]

$$I_{rp} = I_r I_p > 36 \tag{36}.$$

The uncertainty relation that is used to validate the Fisher Information is the Stam, Cramer-Rao inequalities [49, 50]. For Fisher information, an increase in is an indication of a higher localization. As such, there exists a decrease in uncertainty and a corresponding increase in the level of accuracy in predicting the particle's localization. For Shannon entropy, the lower the entropic value, the higher the particle localization [48].

## 6. Results and Discussion

In this study, we first obtained the analytical expressions of the nonrelativistic energy eigenvalues and their corresponding eigenfunctions of the SKHP using the NU method. With the help of the energy eigenvalues of Eq. (7), we generate the numerical results of the energy eigenvalues of SKHP for various potential parameters and quantum states, as shown in Table 1. At each quantum state, the bound state energy eigenvalues increase as the screening parameter increases. The potential parameters of Eq. (1) have also been altered to obtain some special cases. These include the analytical relations for screened Kratzer potential and Hellmann potential. Our results for these special cases agree with the results obtained from the literature. The numerical results of the Hellmann potential are presented in Table 2. These results are consistent with other results obtained by other authors (see Table 2).

We employed the nonrelativistic energy eigenvalues of Eq. (7) to evaluate the rotational-vibrational partition function of the SKHP in closed form. With its expression, other thermodynamic functions have been evaluated. The variation of the ro-vibrational partition function and other thermodynamic functions with the temperature parameter are shown in Figs. (1) – (5), for various quantum states. In Fig. (1), we observe a monotonous increase in the partition function as the temperature parameter increases. Also, the increase in the temperature parameter corresponds to the increase in the quantum states considered. The free energy variation with temperature, shown in Fig. (2), exhibits first a sharp increase in energy at a particular temperature. Later, the temperature increases at a particular free energy corresponding to distinct quantum states. In Fig. (3), we observe a slow decrease in mean energy when the temperature is almost constant. As the temperature begins to increase constantly, the mean energy remains constant for each quantum state. The behaviour of the entropy variation with temperature in Fig. (4) is very similar to Fig. (1), where a monotonous increase is observed as the temperature increases for different quantum states considered. Fig. (5) shows the variation of the specific heat capacity with temperature for various quantum states. Here, the specific heat capacity first increases with a slow increase in temperature and then decreases to a constant level, as the temperature increases continuously.

Figures 6 to 10 and tables 3 and 4 are the graphical and numerical representation of SKHP for $n = 0$ and $n = 1$. The potential parameters $V_0 = 3$, $V_1 = 5$, $V_2 = 10$, the reduced mass and Planck constant are in atomic units and the screening parameter, $\alpha$, lies between $0.1$ and $0.9$ were chosen. The Shannon entropy and Fisher information is evaluated for low lying states, $n=0$ and $n=1$ for $l=0$ and $m=0$. In figures 6 and 7 we the probability for ground and first excited states in position and momentum space respectively to study the behavior for the SKHP model across the potential parameter. The difference in behavior here is caused by the concentration of the density distribution in their different coordinate spaces. It should be noted that the analytical derivations of the entropic information are very cumbersome. As such, we refer our readers to Refs. [35-40] for details. Figs. (8) and (9) show the variations the Shannon entropy in position and momentum spaces respectively with the screening parameter $\alpha$, for both ground and first excited states, respectively. The plots show and increasing Shannon entropy in the position space and decrease in the momentum space. It can be deduced that that increasing localization is observed in the momentum space. The numerical results of the Shannon entropies $S_r$ and $S_p$ and their sum $S_T$ for various screening parameters $\alpha$ is presented in table 3. The Shannon entropy showed an increasing trend and turnaround from $\alpha > 0.6$. As it is in the position space, the same was witnessed in the momentum space. Obviously, we see an alternating increasing and decreasing localization across the screening parameter for the different spaces. It is of note that the BBM inequality was satisfied.

Figures 10 and 11 we studied the behavior of the potential model for Fisher information across the screening parameter. Table 3 is the numerical analysis of the Fisher information in position and momentum space. The uncertainty relation is verified in this case as presented in eqn (36). Our result was obtained in the absence of $l$ and $m$, hence, Fisher information becomes proportional to kinetic energy. And so and increasing Fisher would represent an increasing energy and an increasing localization. The momentum space of figure 11 exhibits an increasing localization for the both eigenstates. The position and momentum space witnessed a turnaround in the trend at $\alpha > 0.5$.

## 7. Concluding Remarks

In this work, the radial Schrodinger equation for a newly proposed screened Kratzer-Hellmann potential was studied using the conventional Nikiforov-Uvarov method. The analytical expressions of the energy eigenvalues and normalized eigenfunction were obtained by employing Greene-Aldrich approximation scheme and an appropriate coordinate transformation technique, due to the presence of the centrifugal term encountered. Some special cases were deduced from the energy eigenvalues of the SKHP, by altered the parameters of the new potential. The results of the special cases correspond to the energy eigenvalues of some known potentials, as documented in the literature. We also obtained the numerical results of the energy eigenvalues for SKHP for various quantum states. These results increase with an increase in the quantum states considered.

By employing the energy eigenvalues of the SKHP, we evaluated the rotational-vibrational partition function and other thermodynamic properties of SKHP in closed form. These results have been discussed extensively via the graphical method.

We have study the Shannon entropy and Fisher information of SKHP for ground and first excited states in both position and momentum spaces numerically. Our results for both states agree with the BBM inequality for Shannon entropy and Stam, Crammer-Rao inequality for Fisher information. Our result shows an alternating increasing and decreasing localization across the potential parameters

This research can equally be extended to the other physical systems including different diatomic molecules, using our newly proposed potential model and also for other global theoretic measures as such Renyi entropy, Tsallis entropy and Onicescu energy.

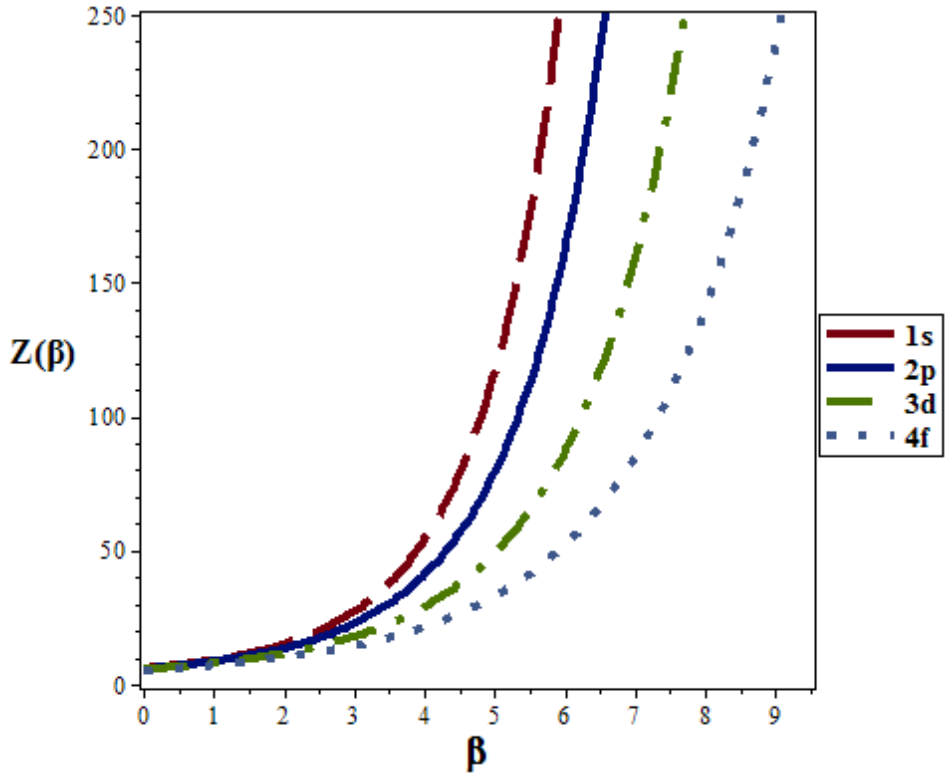

Fig. 1: Partition function versus $\beta$ for various quantum states.

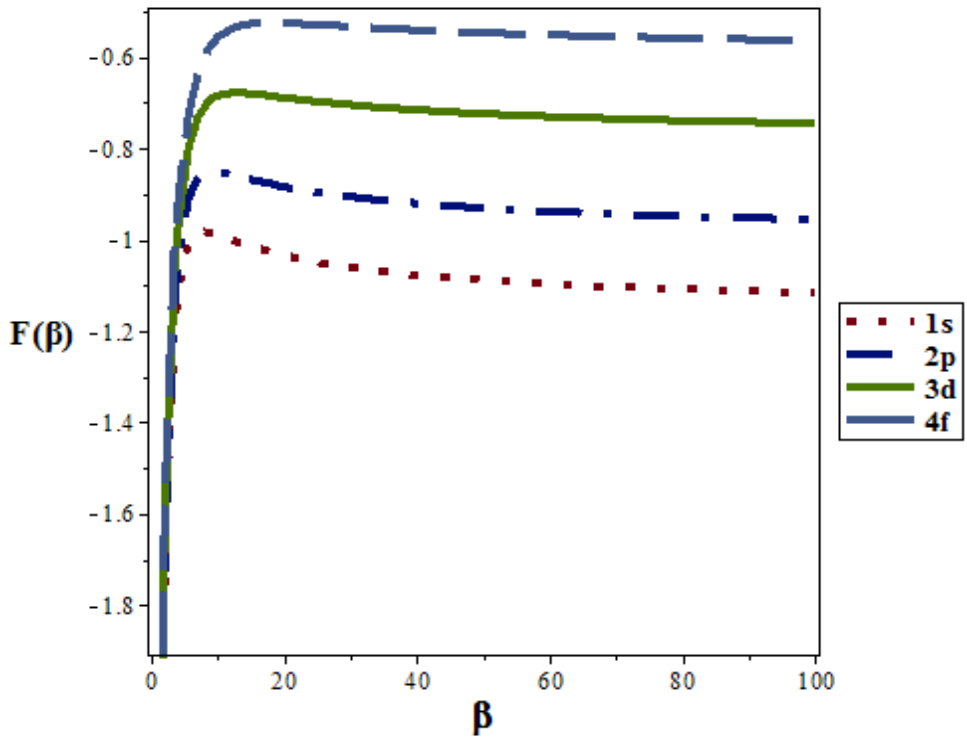

Fig. 2: Free energy versus $\beta$ for various quantum states.

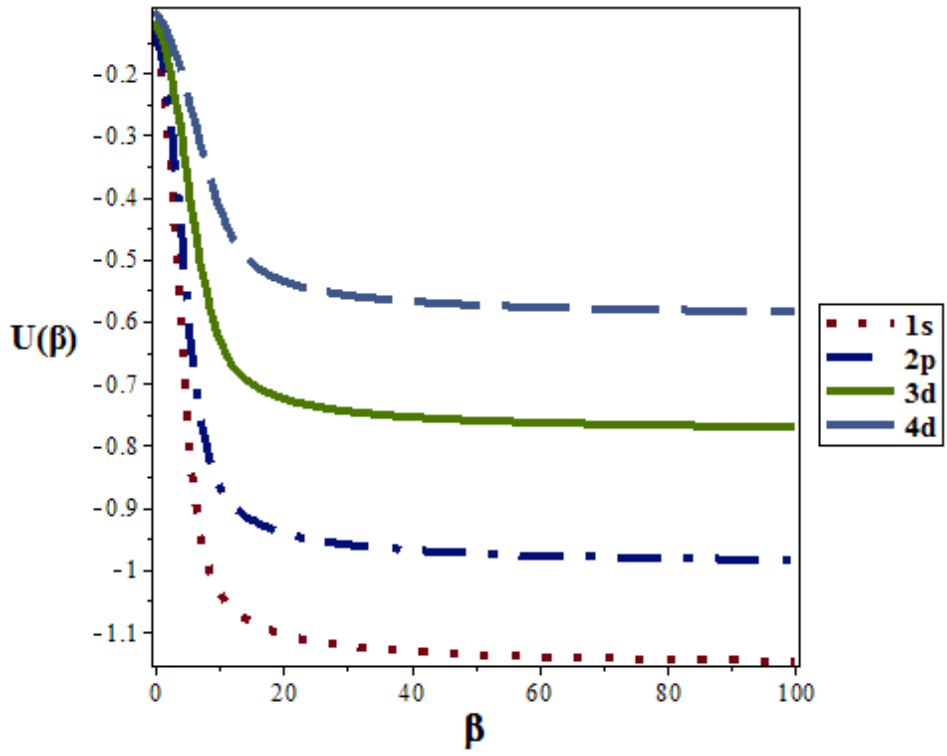

Fig. 3: Mean energy versus $\beta$ for various quantum states.

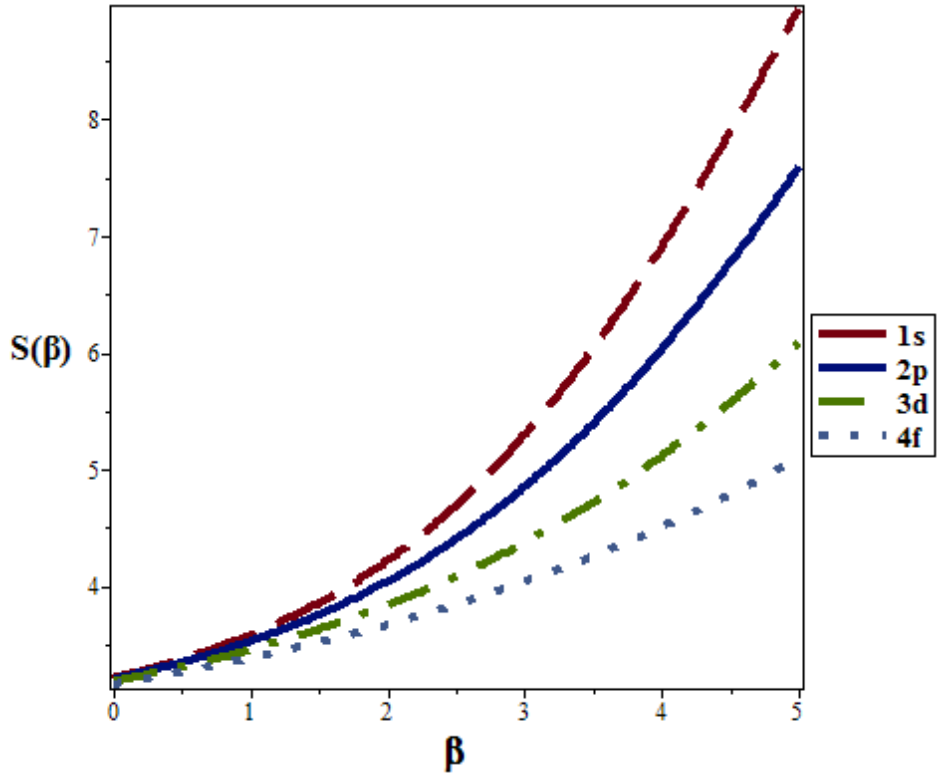

Fig. 4: Entropy versus $\beta$ for various quantum states.

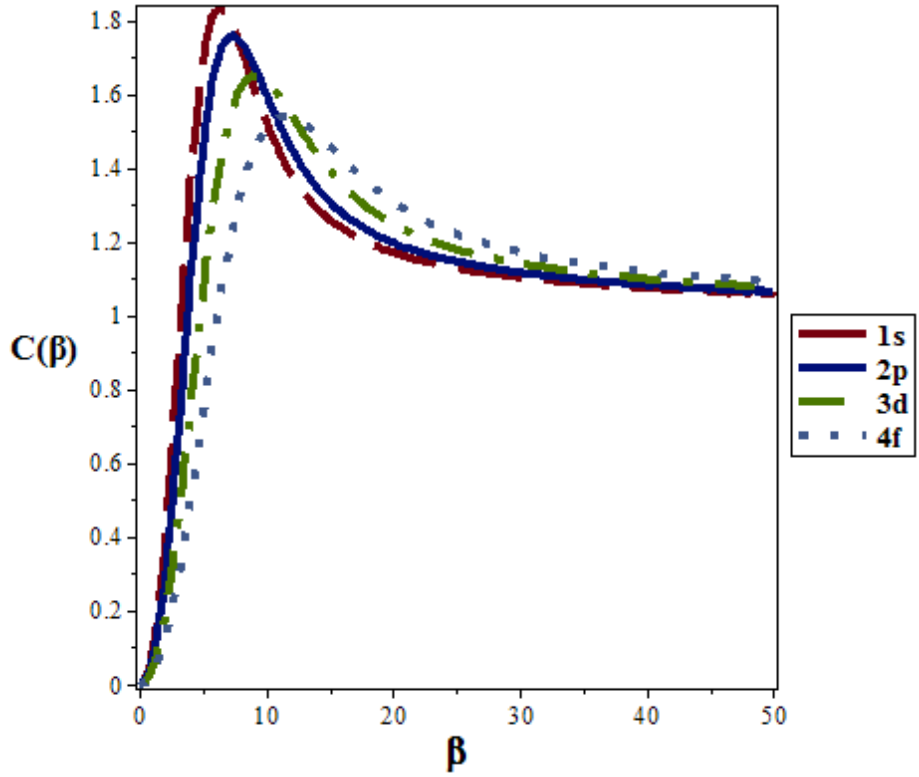

Fig. 5: Specific heat capacity versus $\beta$ for various quantum states.

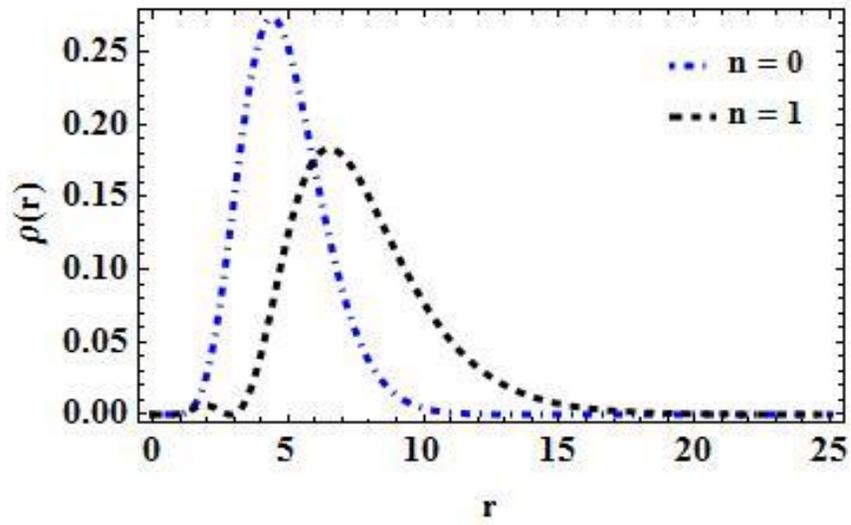

Fig 6: Probability density in position space for ground for $n = 0$ and $n = 1$.

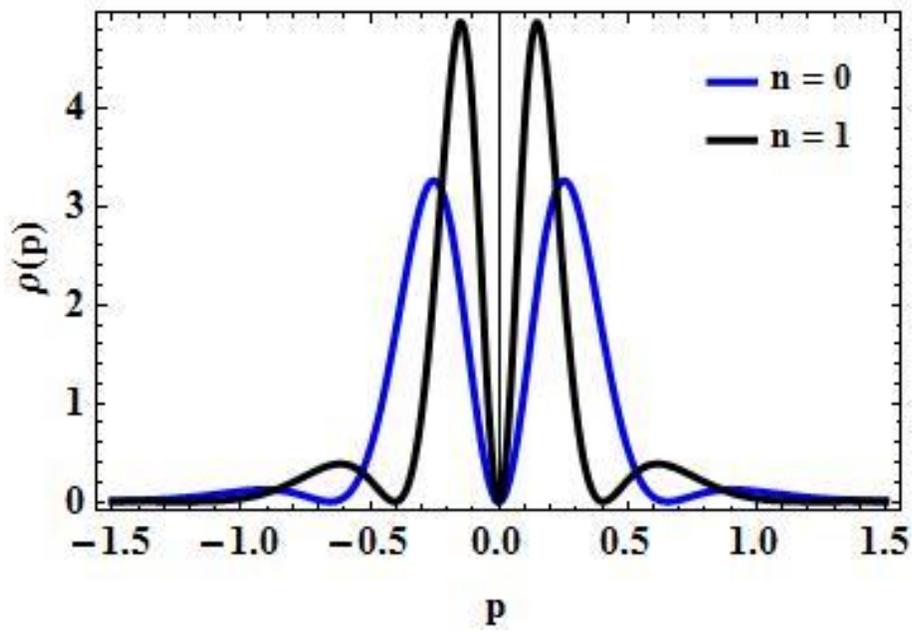

Fig. 7: Probability density in position space for $n = 0$ and $n = 1$.

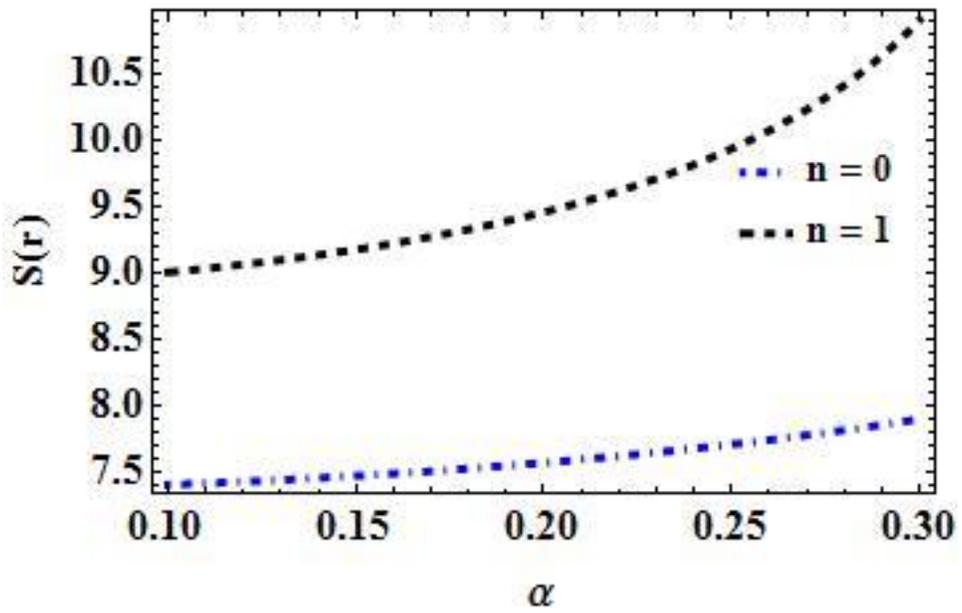

Fig 8: Shannon entropy in position space for for $n = 0$ and $n = 1$.

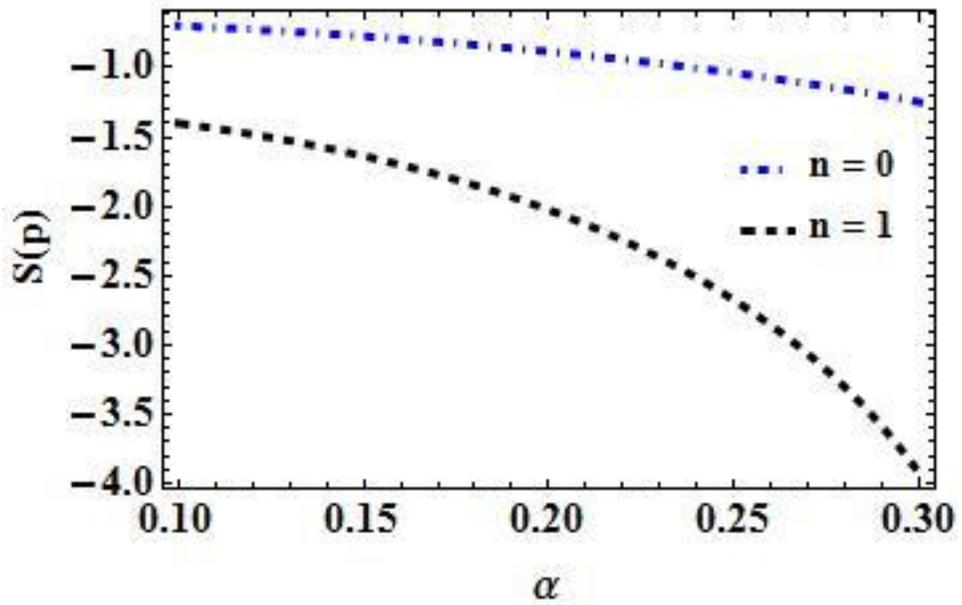

Fig 9: Shannon entropy in momentum space for for $n = 0$ and $n = 1$.

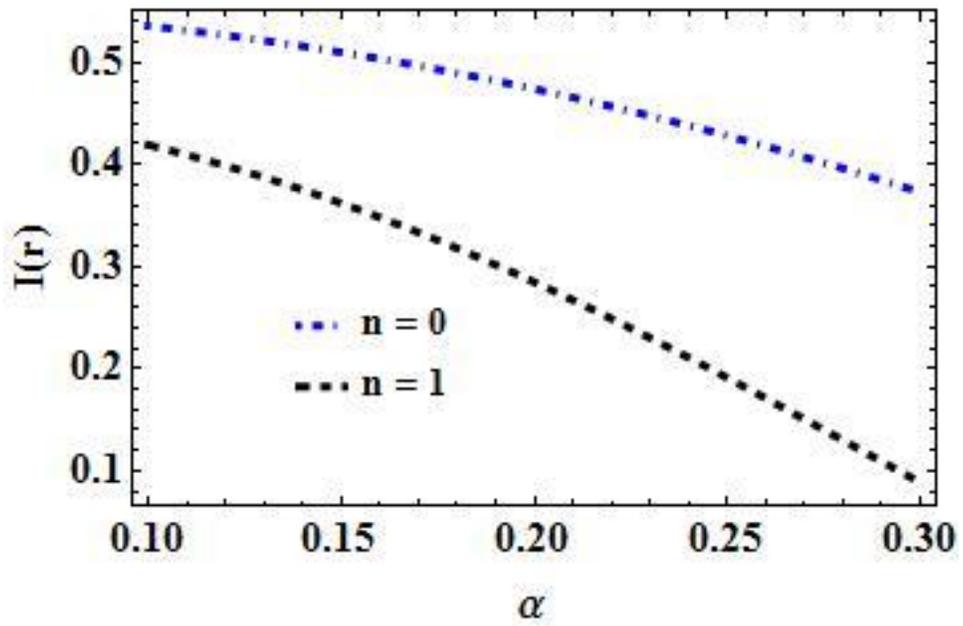

Fig 10: Fisher Information in position space for for $n = 0$ and $n = 1$.

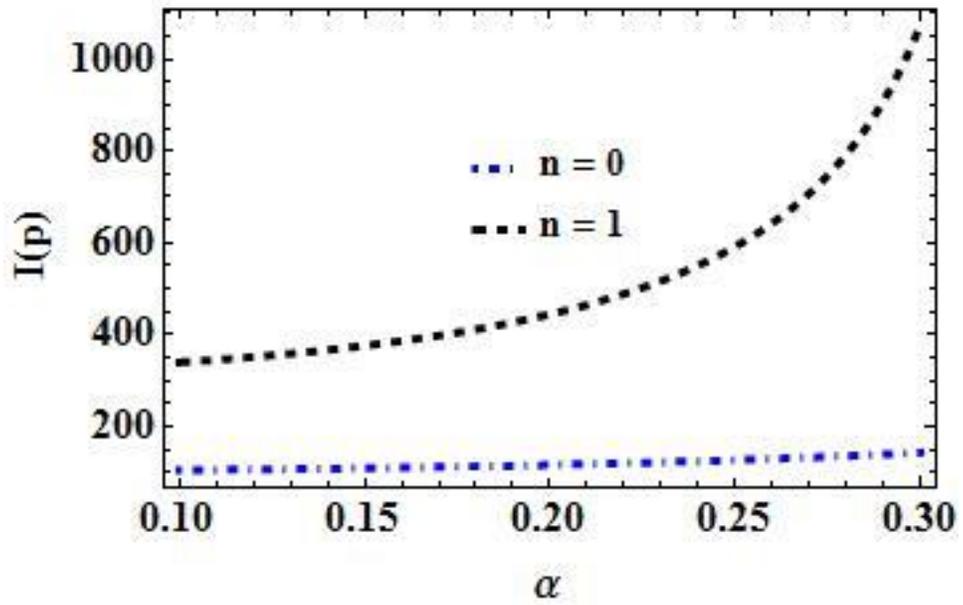

Fig 11: Fisher Information in momentum space for for $n = 0$ and $n = 1$.

Table 1: Energy eigenvalues of Screened Kratzer-Hellmann Potential for different quantum states.

| States | $\alpha$ | $V_0 = 3,\ V_1 = 5,\ V_2 = 10$ | $V_0 = 3,\ V_1 = -5,\ V_2 = 10$ | $V_0 = -3,\ V_1 = 5,\ V_2 = 10$ |
|--------|---------|------------------------------|-------------------------------|--------------------------------|
| 1s | 0.001 | −1.166753487 | −0.07698780420 | −0.06898780425 |
|    | 0.005 | −1.162835696 | −0.09307001360 | −0.05307001360 |
|    | 0.01  | −1.158092602 | −0.1133269179  | −0.03332691790 |
|    |       |              |                |                |
| 2s | 0.001 | −0.7228339375 | −0.04924480187 | −0.04124480190 |
|    | 0.005 | −0.7189665660 | −0.06537742998 | −0.02537743000 |
|    | 0.01  | −0.7143810285 | −0.08579189286 | −0.005791892900 |
|    |       |              |                |                |
| 2p | 0.001 | −0.9995030620 | −0.06637806250 | −0.05862806250 |
|    | 0.005 | −0.9975765620 | −0.08195156250 | −0.04320156250 |
|    | 0.01  | −0.9953062500 | −0.1015562500  | −0.02405625000 |
|    |       |              |                |                |
| 3s | 0.001 | −0.4911975333 | −0.03476996481 | −0.02676996484 |
|    | 0.005 | −0.4873925800 | −0.05096501170 | −0.01096501170 |
|    | 0.01  | −0.4830021015 | −0.07157453314 | 0.008425466860 |
|    |       |              |                |                |
| 3p | 0.001 | −0.6393252900 | −0.04392529000 | −0.03608529000 |
|    | 0.005 | −0.6367322500 | −0.05973225000 | −0.02053225000 |
|    | 0.01  | −0.6337290000 | −0.07972900000 | −0.001329000000 |
|    |       |              |                |                |
| 3d | 0.001 | −0.7794759730 | −0.05241685984 | −0.04500133815 |
|    | 0.005 | −0.7802135070 | −0.06730961050 | −0.03023200220 |
|    | 0.01  | −0.7812511260 | −0.08604125025 | −0.01188603365 |
|    |       |              |                |                |
| 4s | 0.001 | −0.3552719036 | −0.02627751996 | −0.01827751998 |
|    | 0.005 | −0.3515413695 | −0.04254698558 | −0.002546985580 |
|    | 0.01  | −0.3473834492 | −0.06338906561 | 0.01661093440 |
|    |       |              |                |                |
| 4p | 0.001 | −0.4436746944 | −0.03173025000 | −0.02384136110 |
|    | 0.005 | −0.4407562500 | −0.04770069445 | −0.008256250000 |
|    | 0.01  | −0.4374694444 | −0.06802500000 | 0.01086388890 |
|    |       |              |                |                |
| 4d | 0.001 | −0.5227782730 | −0.03649562466 | −0.02888786626 |
|    | 0.005 | −0.5220348260 | −0.05182976205 | −0.01379097022 |
|    | 0.01  | −0.5213279325 | −0.07121984755 | 0.004857736110 |
|    |       |              |                |                |
| 4f | 0.001 | −0.5886338120 | −0.04030298094 | −0.03318478305 |
|    | 0.005 | −0.5917392760 | −0.05459042350 | −0.01899943412 |
|    | 0.01  | −0.5957168180 | −0.07254543952 | −0.001363460750 |

Table 2: Energy eigenvalues of Hellmann Potential ( $V_0 = -1$, $V_1 = -2$, $V_2 = 0$ ) for different quantum states.

| States | $\alpha$ | Present | SUSY [*5] | NU [1] | AP [1] |
|--------|-------|-------------|-----------|-----------|-----------|
| 1s | 0.001 | −2.250500250 | −2.250500 | −2.250500 | −2.248981 |
|    | 0.005 | −2.252506250 | −2.252510 | −2.252506 | −2.244993 |
|    | 0.01  | −2.255025000 | −2.255020 | −2.255025 | −2.240030 |
| 2s | 0.001 | −0.5630010000 | −0.563750 | −0.563001 | −0.561502 |
|    | 0.005 | −0.5650250000 | −0.568756 | −0.565025 | −0.557549 |
|    | 0.01  | −0.5676000000 | −0.575025 | −0.567600 | −0.552697 |
| 2p | 0.001 | −0.5622502500 | −0.562999 | −0.563000 | −0.561502 |
|    | 0.005 | −0.5612562500 | −0.564975 | −0.565000 | −0.557541 |
|    | 0.01  | −0.5600250000 | −0.567400 | −0.567500 | −0.552664 |
| 3s | 0.001 | −0.2505022500 | −0.251500 | −0.250502 | −0.249004 |
|    | 0.005 | −0.2525562500 | −0.257506 | −0.252556 | −0.245110 |
|    | 0.01  | −0.2552250000 | −0.265025 | −0.255225 | −0.240435 |
| 3p | 0.001 | −0.2501680278 | −0.251165 | −0.250501 | −0.249004 |
|    | 0.005 | −0.2508673611 | −0.255801 | −0.252531 | −0.245102 |
|    | 0.01  | −0.2518027778 | −0.261536 | −0.255125 | −0.240404 |
| 3d | 0.001 | −0.2495002500 | −0.250496 | −0.250833 | −0.249003 |
|    | 0.005 | −0.2475062500 | −0.252406 | −0.254151 | −0.245086 |
|    | 0.01  | −0.2450250000 | −0.254625 | −0.258269 | −0.240341 |
| 4s | 0.001 | −0.1411290000 | −0.142250 | −0.141129 | −0.139633 |
|    | 0.005 | −0.1432250000 | −0.148756 | −0.143225 | −0.135819 |
|    | 0.01  | −0.1460250000 | −0.156900 | −0.146025 | −0.131380 |
| 4p | 0.001 | −0.1409405625 | −0.142061 | −0.141128 | −0.139632 |
|    | 0.005 | −0.1422640625 | −0.147777 | −0.143200 | −0.135811 |
|    | 0.01  | −0.1440562500 | −0.154856 | −0.145925 | −0.131350 |
| 4d | 0.001 | −0.1405640625 | −0.141683 | −0.141314 | −0.139632 |
|    | 0.005 | −0.1403515625 | −0.145827 | −0.144089 | −0.135795 |
|    | 0.01  | −0.1401562500 | −0.150806 | −0.147606 | −0.131290 |
| 4f | 0.001 | −0.1400002500 | −0.141117 | −0.141686 | −0.139631 |
|    | 0.005 | −0.1375062500 | −0.142925 | −0.145902 | −0.135772 |
|    | 0.01  | −0.1344000000 | −0.144825 | −0.151106 | −0.131200 |

TABLE 3: Numerical results for Shannon entropies for the SKHP for various values of $\alpha$.

| $\alpha$ | $S_r(n=0)$ | $S_p(n=0)$ | $S_t(n=0)$ | $S_r(n=1)$ | $S_p(n=1)$ | $S_t(n=1)$ |
|---|---|---|---|---|---|---|
| 0.1 | 7.41368 | -0.698858 | 6.71483 | 9.01343 | -1.40064 | 7.61279 |
| 0.2 | 7.57899 | -0.883998 | 6.695 | 9.46499 | -2.02577 | 7.43922 |
| 0.3 | 7.90697 | -1.24789 | 6.65908 | 10.9093 | -3.92553 | 6.98378 |
| 0.4 | 8.54451 | -1.93976 | 6.60475 | 11.7684 | -5.02956 | 6.73885 |
| 0.5 | 10.1283 | -3.57873 | 6.54958 | 8.794 | -1.70107 | 7.09292 |
| 0.6 | 14.6895 | -8.12546 | 6.564 | 7.5468 | -0.32121 | 7.22559 |
| 0.7 | 9.14613 | -2.59677 | 6.54936 | 6.70868 | 0.583515 | 7.2922 |
| 0.8 | 7.60695 | -1.03916 | 6.56779 | 6.06743 | 1.26452 | 7.33196 |
| 0.9 | 6.68032 | -0.0957764 | 6.58454 | 5.54501 | 1.81329 | 7.3583 |

TABLE 4: Numerical results for Fisher Information for the SKHP for various values of $\alpha$.

| $\alpha$ | $I_r(n=0)$ | $I_p(n=0)$ | $I_r I_p(n=0)$ | $I_r(n=1)$ | $I_p(n=1)$ | $I_r I_p(n=1)$ |
|---|---|---|---|---|---|---|
| 0.1 | 0.535186 | 104.894 | 56.138 | 0.41877 | 339.761 | 142.282 |
| 0.2 | 0.473564 | 116.398 | 55.1219 | 0.284545 | 444.495 | 126.479 |
| 0.3 | 0.37255 | 143.263 | 53.3726 | 0.0885678 | 1074.82 | 95.1949 |
| 0.4 | 0.236445 | 215.858 | 51.0384 | 0.0449838 | 1872.04 | 84.2114 |
| 0.5 | 0.0808561 | 620.554 | 50.1756 | 0.380109 | 266.79 | 101.409 |
| 0.6 | 0.00388482 | 13525.4 | 52.5435 | 0.925548 | 118.897 | 110.045 |
| 0.7 | 0.155645 | 322.496 | 50.195 | 1.66802 | 68.8408 | 114.828 |
| 0.8 | 0.434823 | 114.888 | 49.9561 | 2.60518 | 45.2285 | 117.828 |
| 0.9 | 0.811615 | 62.0363 | 50.3496} | 3.73624 | 32.085 | 119.877 |

**APPENDIX**

## A. THE NIKIFOROV-UVAROV METHOD

The Nikiforov-Uvarov (NU) method is based on solving the hypergeometric-type second-order differential equations through the special orthogonal functions [28]. The main equation which is closely associated with the method is given in the following form

$$\psi''(s) + \frac{\tilde{\tau}(s)}{\sigma(s)}\psi'(s) + \frac{\tilde{\sigma}(s)}{\sigma^2(s)}\psi(s) = 0 \tag{A1},$$

where $\sigma(s)$ an $\tilde{\sigma}(s)$ are polynomials at most second-degree, $\tilde{\tau}(s)$ is a first-degree polynomial and $\psi(s)$ is a function of the hypergeometric-type.

The exact solution of Eq. (A1) can be obtained by using the transformation

$$\psi(s) = \phi(s)\chi_n(s) \tag{A2}.$$

This transformation reduces Eq. (A1) into a hypergeometric-type equation of the form

$$\sigma(s)\chi_n''(s) + \tau(s)\chi_n'(s) + \lambda\chi_n(s) = 0 \tag{A3}.$$

The function $\phi(s)$ can be defined as the logarithm derivative

$$\frac{\phi'(s)}{\phi(s)} = \frac{\pi(s)}{\sigma(s)} \tag{A4},$$

where

$$\pi(s) = \frac{1}{2}\left[\tau(s) - \tilde{\tau}(s)\right] \tag{A5},$$

with $\pi(s)$ being at most a first-degree polynomial. The term $\chi_n(s)$ in Eq. (A3) is the hypergeometric function with its polynomial solution given by Rodrigues relation:

$$\chi_n^{(n)}(s) = \frac{B_n}{\rho(s)} \frac{d^n}{ds^n} \left[ \sigma^n \rho(s) \right] \tag{A6}.$$

In the above relation, $B_n$ is the normalization constant and $\rho(s)$ is the weight function which must satisfy the condition:

$$\left( \sigma(s) \rho(s) \right)' = \sigma(s) \tau(s) \tag{A7}$$

and

$$\tau(s) = \tilde{\tau}(s) + 2\pi(s) \tag{A8}.$$

It should be noted that the derivative of $\tau(s)$ with respect to $s$ should be negative. The eigenfunctions and eigenvalues can be obtained using the definition of the function, $\pi(s)$ and the parameter $\lambda$, given respectively as:

$$\pi(s) = \frac{\sigma'(s) - \tilde{\tau}(s)}{2} \pm \sqrt{\left( \frac{\sigma'(s) - \tilde{\tau}(s)}{2} \right)^2 - \tilde{\sigma}(s) + k\sigma(s)} \tag{A9},$$

where

$$k = \lambda - \pi'(s) \tag{A10}.$$

The value of $k$ can be obtained by setting the discriminant of the square root in Eq. (A9) equal to zero. Thus, the new eigenvalue equation can be given as

$$\lambda_n = -n\tau'(s) - \frac{n(n-1)}{2} \sigma''(s), \quad n = 0, 1, 2, \dots \tag{A11}.$$